\documentclass[prd,aps,onecolumn,floatfix,superscriptaddress,nofootinbib,preprintnumbers]{revtex4}
\usepackage{hyperref}

\usepackage{graphics,graphicx}

\usepackage{mathbbol}
\usepackage{amsmath}

\usepackage{color}
\usepackage[normalem]{ulem}

\allowdisplaybreaks

\begin{document}

\preprint{JLAB-THY-25-4235}

\title{\bf Comment on ``QCD factorization with multihadron fragmentation functions''}

\newcommand*{\LVC}{Department of Physics, Lebanon Valley College, Annville, Pennsylvania 17003, USA}\affiliation{\LVC}
\newcommand*{\WM}{Department of Physics, William and Mary, Williamsburg, Virginia 23185, USA}\affiliation{\WM}
\newcommand*{\TU}{Department of Physics, SERC, Temple University, Philadelphia, Pennsylvania 19122, USA}\affiliation{\TU}
\newcommand*{\PSU}{Division of Science, Penn State University Berks, Reading, Pennsylvania 19610, USA}\affiliation{\PSU}
\newcommand*{\JLAB}{Jefferson Lab, Newport News, Virginia 23606, USA}\affiliation{\JLAB}

\author{D.~Pitonyak}\affiliation{\LVC}
\author{C.~Cocuzza}\affiliation{\WM}
\author{A.~Metz}\affiliation{\TU}
\author{A.~Prokudin}\affiliation{\PSU}\affiliation{\JLAB}
\author{N.~Sato}\affiliation{\JLAB}

\begin{abstract}
\noindent We make several comments on the recent work in Ref.~\cite{Rogers:2024nhb} while also reaffirming and adding to the work in Ref.~\cite{Pitonyak:2023gjx}.  We show that the factorization formula for $e^+e^-\to (h_1\cdots h_n)\, X$ in Ref.~\cite{Rogers:2024nhb} is equivalent to a version one can derive using the definition of a $n$-hadron fragmentation function (FF) introduced in Ref.~\cite{Pitonyak:2023gjx}. 
In addition, we scrutinize how to generalize the number density definition of a single-hadron FF to a $n$-hadron FF, arguing that the definition given in Ref.~\cite{Pitonyak:2023gjx} should be considered the standard one.  We also emphasize that the evolution equations for dihadron FFs~(DiFFs) in Ref.~\cite{Pitonyak:2023gjx} have the same splitting functions as those for single-hadron FFs. 
Therefore, the DiFF (and $n$-hadron FF) definitions in Ref.~\cite{Pitonyak:2023gjx} have a natural number density interpretation and are consistent with collinear factorization using the standard hard factors and evolution kernels.
Moreover, we make clear that the operator definition for the DiFF $D_1^{h_1h_2}(\xi,M_h)$ written down in Ref.~\cite{Rogers:2024nhb} agrees exactly with the one in Ref.~\cite{Pitonyak:2023gjx}.  
Contrary to what is implied in Ref.~\cite{Rogers:2024nhb}, this definition did not appear in the literature prior to the work in Ref.~\cite{Pitonyak:2023gjx}.   There also seem to be inconsistencies in how $D_1^{h_1h_2}(\xi,M_h)$ appears in previous unpolarized cross section formulas in the literature. 
\end{abstract}

\maketitle 

%%%%%%%%%%%%%%%%%%%%%%%%%%%%%%%%%%%%%%%%%%
\section{Introduction}
%%%%%%%%%%%%%%%%%%%%%%%%%%%%%%%%%%%%%%%%%%
The dynamics of quarks and gluons (partons) hadronizing in the final state of a high-energy collision can be encoded in non-perturbative objects called fragmentation functions~(FFs)~\cite{Collins:1981uw,Metz:2016swz}, where there has been renewed interest in the study of multi-hadron FFs~\cite{Pitonyak:2023gjx,Rogers:2024nhb}.  Fragmentation functions are key ingredients to factorization theorems of measurable cross sections and are universal across different processes. They
are defined in terms of a quantum field-theoretic operator and have the physical interpretation as number densities for finding a hadron in a parton.  
For single-hadron FFs this can be demonstrated through a parton model calculation of the cross section for $e^+e^-\to hX$ and derivation of charge (or number) and momentum sum rules~\cite{Collins:1981uw,Collins:2011zzd}.  In particular, the standard single-hadron FF $D^{h}_1(\xi, \vec{P}_\perp)$ is a number density in the lightcone momentum fraction $\xi$ of the parton carried by the hadron and the transverse momentum $\vec{P}_\perp$ of the hadron w.r.t.~the parton.  Alternatively, one can change to a frame where the hadron has no transverse momentum and the parton has a transverse momentum $\vec{k}_T$ (and the parton has the same large minus-lightcone  momentum $k^-$).  In this case, $\vec{P}_\perp=-\xi\vec{k}_T$, and one may choose to write the FF as $D_1^{h}(\xi, - \xi \vec{k}_T)$. 

For a dihadron FF (DiFF), where two hadrons are detected together in the final state, the situation is more complicated, since now it is a six-dimensional function. 
Let us emphasize at this point that currently there is no proof of factorization to all orders in a process like $e^+e^-\to (h_1 h_2)\,X$ that involves both DiFFs and single-hadron FFs.  However, the question of how, for any given set of variables, to define a DiFF so that it is a number density and consistent with a parton model factorized expression was  
addressed in detail for the first time in Ref.~\cite{Pitonyak:2023gjx}. Establishing the correct  quantum field-theoretic operator definition for DiFFs (and $n$-hadron FFs), and showing physical observables can be consistently factorized in terms of it, is critical
if one is to fully explore the rich structure provided by multi-hadron dynamics in high-energy collisions. 
A natural starting point is to define a DiFF $D_1^{h_1h_2}(\xi_1,\xi_2,\vec{P}_{1\perp},\vec{P}_{2\perp})$\footnote{We change notation slightly from Ref.~\cite{Pitonyak:2023gjx} for clarity, where we now explicitly write as the arguments of the FF in which variables it is a number density, rather than the different scalar variable combinations it can depend on.} that is a number density in the lightcone momentum fractions $(\xi_1,\xi_2)$ carried by each hadron and their transverse momenta $(\vec{P}_{1\perp},\vec{P}_{2\perp})$ relative to the parton.\footnote{As pointed out in Ref.~\cite{Pitonyak:2023gjx}, the previous definition of (fully unintegrated) DiFFs in the literature~\cite{Bianconi:1999cd} is not a number density in those variables.} A number density for any other set of variables is then defined by using Eq.~(13) of Ref.~\cite{Pitonyak:2023gjx} to account for the Jacobian of the variable transformation from $(\xi_1,\xi_2,\vec{P}_{1\perp},\vec{P}_{2\perp})$ to the new set of variables.  The claims of a number density interpretation in Ref.~\cite{Pitonyak:2023gjx} were  supported through parton model calculations of cross sections in $e^+e^-\to (h_1 h_2)\,X$  and derivations of number and momentum sum rules.   
While recently sum rules for FFs have been questioned~\cite{Collins:2023cuo}, we emphasize that parton model calculations and ``na\"{i}ve'' sum rules are what confirm the accepted number density operator definition of single-hadron FFs~\cite{Collins:1981uw,Collins:2011zzd}.  The approach in Ref.~\cite{Pitonyak:2023gjx} for DiFFs is exactly in the same vein. There was also a generalization given in Eq.~(10) of Ref.~\cite{Pitonyak:2023gjx} for how a $n$-hadron FF $D_1^{\{h_i\}_n}(\{\xi_i\}_n,\{\vec{P}_{i\perp}\}_n)$ can be defined as a number density in $(\xi_1,\dots,\xi_n, \vec{P}_{1\perp},\dots,\vec{P}_{n\perp})$.  (We are using the shorthand notation $\{h_i\}_n\equiv h_1\cdots h_n$, $\{\xi_i\}_n\equiv \xi_1,\dots,\xi_n$, $\{\vec{P}_{i\perp}\}_n\equiv \vec{P}_{1\perp},\dots,\vec{P}_{n\perp}$.)

In Ref.~\cite{Rogers:2024nhb}, a factorization formula is put forth in $e^+e^-$ annihilation for the production of a small-mass cluster of $n$ hadrons, $e^+e^-\to (h_1\cdots h_n)\, X$, in terms of a $n$-hadron FF $d(\xi,-\xi\vec{k}_T,\{P_h\})$  whose operator definition differs from $D_1^{\{h_i\}_n}(\{\xi_i\}_n,\{\vec{P}_{i\perp}\}_n)$ in Ref.~\cite{Pitonyak:2023gjx}.  The derivation in Ref.~\cite{Rogers:2024nhb}  is  a leading-order (parton model) calculation that is promoted to a general factorization formula.  No higher-order calculation was performed.  
Consequently, when we use the word ``factorization'' in this paper, we are doing so to align with the language of Ref.~\cite{Rogers:2024nhb}, not as any statement or confirmation of all-orders factorization for $e^+e^-\to (h_1\cdots h_n)\, X$. With regard to the fundamental definition of a $n$-hadron FF as a number density, we emphasize that factorization cannot give a different definition than one based on parton model calculations and sum rule derivations (as performed in Ref.~\cite{Pitonyak:2023gjx}).

In the following, we make several comments on Ref.~\cite{Rogers:2024nhb} while reaffirming the work of Ref.~\cite{Pitonyak:2023gjx} and adding some new calculations that further support its results.  After setting up some notation and definitions in Sec.~\ref{s:not}, we show in Sec.~\ref{s:fac} that the factorization formula in Ref.~\cite{Rogers:2024nhb} involving $d(\xi,-\xi\vec{k}_T,\{P_h\})$ is equivalent to one involving $D_1^{\{h_i\}_n}(\{\xi_i\}_n,\{\vec{P}_{i\perp}\}_n)$ with the same standard hard factor, and we show explicitly how to connect the two.  In Sec.~\ref{s:num} we examine the number density interpretations of $d(\xi,-\xi\vec{k}_T,\{P_h\})$ and $D_1^{\{h_i\}_n}(\{\xi_i\}_n,\{\vec{P}_{i\perp}\}_n)$, arguing that for the latter it is unambiguous and a natural starting point from which one can define $n$-hadron FFs as number densities in other variables.  In Sec.~\ref{s:evo} we briefly discuss the evolution equations for DiFFs derived in Ref.~\cite{Pitonyak:2023gjx}, emphasizing that they have the same standard splitting functions as single-hadron FFs.  In Sec.~\ref{s:prevdef} we focus on the operator definition of $D_1^{h_1h_2}(\xi,M_h)$ that is especially important for phenomenology.  We make clear that what Ref.~\cite{Rogers:2024nhb} uses for the operator definition of $D_1^{h_1h_2}(\xi,M_h)$ is exactly the same as what was already introduced in Ref.~\cite{Pitonyak:2023gjx}, and it did not appear in the literature before that. There also seem to be inconsistencies in how $D_1^{h_1h_2}(\xi,M_h)$ appears in previous unpolarized cross section formulas in the literature.  We summarize in~Sec.~\ref{s:concl}.

%%%%%%%%%%%%%%%%%%%%%%%%%%%%%%%%%%%%%%%%%%
\section{Notation and Fragmentation Function Definitions\label{s:not}}
%%%%%%%%%%%%%%%%%%%%%%%%%%%%%%%%%%%%%%%%%%
We refer the reader to Refs.~\cite{Rogers:2024nhb, Pitonyak:2023gjx} for a detailed discussion of different reference frames and kinematic variables.\footnote{In Ref.~\cite{Rogers:2024nhb}, the fragmenting parton has a large plus-lightcone momentum $k^+$, whereas here we will follow Ref.~\cite{Pitonyak:2023gjx} and take it to have a large minus-lightcone momentum $k^-$.}  We mention here that the hadrons carry individual momentum fractions $\xi_i = P_i^-/k^-$ and have a total momentum fraction $\xi\equiv\sum_{i=1}^n \xi_i$, where $k$ is the momentum of the fragmenting parton, $P_i$ is the momentum of hadron $h_i$, and $P_h\equiv\sum_{i=1}^n \!P_i$ with $P_h^2=M_h^2$.  
For the case of a dihadron, we also have $\zeta\equiv(P_1^--P_2^-)/P_h^-=(\xi_1-\xi_2)/\xi$ and $\vec{R}_T \equiv (\vec{P}_{1T}-\vec{P}_{2T})/2$.\footnote{As in Ref.~\cite{Pitonyak:2023gjx}, we will use $\perp (T)$ to denote transverse components in the parton ($n$-hadron) frame.}  We start with the following $n$-hadron fragmentation correlator (which extends the dihadron correlator from Ref.~\cite{Bianconi:1999cd} to include $n$ hadrons in the final state),
\begin{equation}
    \Delta^{\{h_i\}_n}\equiv{\rm Tr}\sum_X\hspace{-0.5cm}\int\! \int\!\frac{dx^+d^2\vec{x}_\perp}{(2\pi)^3}\,e^{ik\cdot x}\langle 0|\gamma^-\psi(x)|P_1,\dots,P_n;X\rangle\langle P_1,\dots,P_n;X|\bar{\psi}(0)|0\rangle\big|_{x^-=0}\,.\label{e:DiFF_corr} 
\end{equation}
This correlator is used in both Ref.~\cite{Rogers:2024nhb} and Ref.~\cite{Pitonyak:2023gjx} as the foundation through which $n$-hadron FFs are defined.   
In Eq.~(10) of Ref.~\cite{Pitonyak:2023gjx}, a fundamental $n$-hadron FF $D_1^{\{h_i\}_n}(\{\xi_i\}_n,\{\vec{P}_{i\perp}\}_n)$ is introduced that is a number density in the $3n$ variables $(\xi_1,\dots,\xi_n, \vec{P}_{1\perp},\dots,\vec{P}_{n\perp})$,
\begin{equation}
    D_1^{\{h_i\}_n}(\{\xi_i\}_n,\{\vec{P}_{i\perp}\}_n)=\frac{1}{4(16\pi^3)^{n-1}\xi_1\cdots\xi_n}\,\Delta^{\{h_i\}_n}\,, \label{e:D1hn}
\end{equation}
where the prefactor in front is needed in order to be able to use the number or momentum operator for each hadron in sum rule derivations (see Supplemental Material Secs.~S1 and S2 of Ref.~\cite{Pitonyak:2023gjx} and also Sec.~\ref{s:num} below). A number density for any set of variables is then defined by using Eq.~(13) of Ref.~\cite{Pitonyak:2023gjx} to account for the Jacobian of the variable transformation.

In Eq.~(79) of Ref.~\cite{Rogers:2024nhb}, a different operator definition of a $n$-hadron FF, denoted $d(\xi,-\xi\vec{k}_T,\{P_h\})$, was put forth, 
\begin{equation}
    d(\xi,-\xi\vec{k}_T,\{P_h\}) = \frac{1}{4\xi}\,\Delta^{\{h_i\}_n}\,,\label{e:D1hn_79}
\end{equation}
which differs from Eq.~(\ref{e:D1hn}) above in the prefactor and arises from the reasoning that the factorization steps for $e^+e^-\to (h_1\cdots h_n)\, X$ do not change from the single-hadron case~\cite{Rogers:2024nhb}.  

In the following sections we will elaborate on the differences between $D_1^{\{h_i\}_n}(\{\xi_i\}_n,\{\vec{P}_{i\perp}\}_n)$ from Ref.~\cite{Pitonyak:2023gjx} and $d(\xi,-\xi\vec{k}_T,\{P_h\})$ from Ref.~\cite{Rogers:2024nhb}.
We point out that the argument $\{P_h\}$ in $d(\xi,-\xi\vec{k}_T,\{P_h\})$ is meant to ``symbolize dependence on all of the observed $n$ hadron momenta''~\cite{Rogers:2024nhb}. This is ambiguous and {\it a~priori} prevents one from a meaningful discussion about whether $d(\xi,-\xi\vec{k}_T,\{P_h\})$ represents a number density in a specific set of $3n$ variables. The only definitive statement Ref.~\cite{Rogers:2024nhb} makes in this regard is that, just like the single-hadron FF case that also has the same prefactor as Eq.~(\ref{e:D1hn_79}), $d(\xi,-\xi\vec{k}_T,\{P_h\})$ is a number density in $(\xi, \xi\vec{k}_T)$ after integrating out all the relative internal momenta of the $n$ hadrons.\footnote{Note that factors of $1/(16\pi^3E'_i)$, where $E'_i$ are energy variables in what Ref.~\cite{Rogers:2024nhb} denotes as $dY'$, must also be included when integrating out those relative internal momenta -- see Eqs.~(74), (82), (83) of Ref.~\cite{Rogers:2024nhb}.} However, it is important to know in the first place in which specific variables involving those internal momenta a $n$-hadron FF is a number density. 
A clear number density operator definition at the fully unintegrated level
allows the most information to be gained phenomenologically from the rich structure of multi-hadron dynamics that exist when moving beyond single-hadron fragmentation.

For the $n=2$ (dihadron) case, we explicitly write here for ease of reference the operator definitions for the DiFFs $D^{h_1h_2}_1(\xi,\zeta,\vec{R}_T)$ (Ref.~\cite{Pitonyak:2023gjx}, Eq.~(15)), $D_1^{h_1h_2}(\xi_1,\xi_2,\vec{R}_T)$ (mentioned in the third sentence after Eq.~(16) of Ref.~\cite{Pitonyak:2023gjx}), and $D^{h_1h_2}_1(\xi,M_h)$ (Ref.~\cite{Pitonyak:2023gjx}, Eq.~(17)):
\begin{subequations}
\begin{align}
    D_1^{h_1h_2}(\xi,\zeta,\vec{R}_T) &= \frac{\xi}{32\pi^3(1-\zeta^2)}\int\! d^2\vec{k}_T\,\Delta^{h_1h_2}\,,\label{e:D1zzetaRT}
    \\[0.3cm]
    D_1^{h_1h_2}(\xi_1,\xi_2,\vec{R}_T) &= \frac{\xi^2}{64\pi^3\xi_1\xi_2}\int\! d^2\vec{k}_T\,\Delta^{h_1h_2}\,, \label{e:D1z1z2RT}\\[0.3cm]
     D_1^{h_1h_2}(\xi,M_h) &= \frac{\xi M_h}{64\pi^2}\int \! d\zeta\int\! d^2\vec{k}_T\,\Delta^{h_1h_2}\,.\label{e:D1zMh}
\end{align}
\end{subequations}
These DiFFs are number densities in $(\xi,\zeta,\vec{R}_T)$, $(\xi_1,\xi_2,\vec{R}_T)$, and $(\xi,M_h)$, respectively.\footnote{Since Ref.~\cite{Pitonyak:2023gjx} introduced its $n$-hadron FFs (and DiFFs) independent of any observable, it is clear that the $z$ and $z_i$ variables there correspond to the hadron momentum fractions $\xi$ and $\xi_i$ here (and in Ref.~\cite{Rogers:2024nhb}) and not to external kinematical variables for $e^+e^-\to (h_1h_2)\,X$.  There was no conflation of hadron momentum fractions and external kinematical variables in Ref.~\cite{Pitonyak:2023gjx}.}

%%%%%%%%%%%%%%%%%%%%%%%%%%%%%%%%%%%%%%%%%%
\section{Factorization for $\boldsymbol{e^+e^-\to (h_1\cdots h_n)\, X}$ \label{s:fac}}
%%%%%%%%%%%%%%%%%%%%%%%%%%%%%%%%%%%%%%%%%%
The key argument in Ref.~\cite{Rogers:2024nhb} for the claim that $d(\xi,-\xi\vec{k}_T,\{P_h\})$ should be considered the fundamental definition of a $n$-hadron FF is that it arises in a derivation of factorization for $e^+e^-\to (h_1\cdots h_n)\, X$. We show here that 
$D_1^{\{h_i\}_n}(\{\xi_i\}_n,\{\vec{P}_{i\perp}\}_n)$   also enters a factorization formula for that process.  Some relevant external kinematic variables are $z_i = 2P_i\cdot q/Q^2$, the momentum of the virtual photon $q$, and $Q^2 = q^2$, as well as $z\equiv\sum_{i=1}^n z_i$.  
We readily find Eq.~(3) of Ref.~\cite{Rogers:2024nhb} can equivalently be written as the following factorization formula:
\begin{equation}
    \left({\displaystyle \prod_{i=1}^n} \,\frac{2 E_{i}(2\pi)^3}{d^3\vec{P}_{i}}\right)d\sigma = \frac{(16\pi^3)^{n-1}\,z_1\cdots z_n}{z^2}\int_z^1\!\frac{d\hat{z}}{\hat{z}^{n-1}}\,\left(\frac{2E_{\hat{k}}(2\pi)^3d\hat{\sigma}}{d^3\vec{\hat{k}}}\right)\!\left(\frac{\xi^2}{4(16\pi^3)^{n-1}\xi_1\cdots\xi_n}\!\int \!d^2\vec{k}_{T}\, \Delta^{\{h_i\}_n}\!\right) + {\rm p.s.}\,, \label{e:fact_finala}
\end{equation}
where we used the fact that 
\begin{equation}
    \hat{z} = \frac{z}{\xi} + {\rm p.s.}\,,\quad\quad\frac{z_i}{\hat{z}}=\xi_i+{\rm p.s.}\,,\label{e:zxi}
\end{equation}
with p.s.~indicating ``power-suppressed'' terms $\sim \Lambda^2/Q^2$, where $\Lambda$ denotes a typical hadronic scale.\footnote{Note that Eq.~\eqref{e:zxi} also implies  $(z_1-z_2)/z=\zeta+{\rm p.s.}\,$.}  The factor $2E_{\hat{k}}(2\pi)^3d\hat{\sigma}/d^3\vec{\hat{k}}$ is the usual hard factor for the inclusive production of an on-shell massless parton with 3-momentum $\vec{\hat{k}}$. The last factor in parentheses on the r.h.s.~of Eq.~(\ref{e:fact_finala}) above can be written in terms of the function $D_1^{\{h_i\}_n}(\{\xi_i\}_n,\{\vec{P}_{i\perp}\}_n)$ we defined in Eq.~(\ref{e:D1hn}),
which then brings Eq.~(\ref{e:fact_finala}) into its final form:
\begin{equation}
     \left({\displaystyle \prod_{i=1}^n} \,\frac{2 E_{i}(2\pi)^3}{d^3\vec{P}_{i}}\right)d\sigma = \frac{(16\pi^3)^{n-1}\,z_1\cdots z_n}{z^2}\int_z^1\!\frac{d\hat{z}}{\hat{z}^{n-1}}\,\left(\frac{2E_{\hat{k}}(2\pi)^3d\hat{\sigma}}{d^3\vec{\hat{k}}}\right)\!\left(\xi^2\!\int \!d^2\vec{k}_{T}\,D_1^{\{h_i\}_n}(\{\xi_i\}_n,\{\vec{P}_{i\perp}\}_n)\right) + {\rm p.s.}\,. \label{e:fact_finalb}
\end{equation}
We therefore have a factorization formula involving the $n$-hadron FF from Ref.~\cite{Pitonyak:2023gjx} with the usual hard factor.  We highlight that the measure of the convolution integral in Eq.~(\ref{e:fact_finalb}), which differs from Eq.~(3) of Ref.~\cite{Rogers:2024nhb} (see also Eq.~(\ref{e:fact_3}) below), is consistent with the evolution equations derived in Ref.~\cite{Pitonyak:2023gjx} for the DiFFs -- see Sec.~\ref{s:evo} below. The expression (\ref{e:fact_finalb}) appears in its general form here for the first time, although it follows from the same type of parton model calculations performed in Ref.~\cite{Pitonyak:2023gjx} (as does Eq.~(3) of Ref.~\cite{Rogers:2024nhb}).   

In Sec.~VIII of Ref.~\cite{Rogers:2024nhb}, it is stated that the cross section results from Ref.~\cite{Pitonyak:2023gjx} could not be reproduced. 
The reason is Ref.~\cite{Rogers:2024nhb} did not
correctly identify the DiFFs as we defined them in Eqs.~(\ref{e:D1zzetaRT}), (\ref{e:D1z1z2RT}), (\ref{e:D1zMh}).  For example, what Ref.~\cite{Rogers:2024nhb} calls $d_{\rm mod, red,1}(\xi, M_h)$ in Eq.~(139) is not what was defined as $D_1^{h_1h_2}(\xi,M_h)$ in Ref.~\cite{Pitonyak:2023gjx} (Eq.~(\ref{e:D1zMh}) above).  Indeed, we have verified the original parton model cross section calculations in Ref.~\cite{Pitonyak:2023gjx} for $d\sigma/dz\,dM_h$, $d\sigma/dz_1dz_2d^2\vec{R}_T$, and $d\sigma/dz\,d\zeta\, d^2\vec{R}_T$ starting now from Eq.~(\ref{e:fact_finala}) (for $n=2$).  We also noticed that the result for $d\sigma/dz_1dz_2d^2\vec{R}_T$ (see Eq.~(\ref{e:dsig_genb}) below) using Eq.~(\ref{e:fact_finala}) exactly matches the structure of the ``double fragmentation'' term in the next-to-leading order (NLO) calculation from Ref.~\cite{deFlorian:2003cg} (see also Refs.~\cite{Majumder:2004br,Majumder:2004wh}).

We will now show how Eq.~(\ref{e:fact_finalb}) above and Eq.~(3) of Ref.~\cite{Rogers:2024nhb} are connected.  As argued in Ref.~\cite{Pitonyak:2023gjx} (see also Sec.~\ref{s:num} below), $D_1^{\{h_i\}_n}(\{\xi_i\}_n,\{\vec{P}_{i\perp}\}_n)$ is a number density in $(\xi_1,\dots,\xi_n, \vec{P}_{1\perp},\dots,\vec{P}_{n\perp})$.  In order to define a $n$-hadron FF that is a number density in a different set of variables denoted by $(y_1,\dots,y_n, \vec{Y}_{1},\dots,\vec{Y}_{n})$, we must follow the simple transformation rule in Eq.~(13) of Ref.~\cite{Pitonyak:2023gjx} via the Jacobian,  
\begin{equation}
    D_1^{\{h_i\}_n}(\{y_i\}_n,\{\vec{Y}_{i}\}_n) = D_1^{\{h_i\}_n}(\{\xi_i\}_n,\{\vec{P}_{i\perp}\}_n)\,\left |\frac{\partial(\xi_1,\dots,\xi_n, \vec{P}_{1\perp},\dots,\vec{P}_{n\perp})}{\partial(y_1,\dots,y_n, \vec{Y}_{1},\dots,\vec{Y}_{n})} \right |.\label{e:Jac}
\end{equation}
Recall now Eq.~(3) of Ref.~\cite{Rogers:2024nhb} (using $d\xi/\xi^2 = -d\hat{z}/z + {\rm p.s.}$),
\begin{equation}
     \left({\displaystyle \prod_{i=1}^n} \,\frac{2 E_{i}(2\pi)^3}{d^3\vec{P}_{i}}\right)d\sigma = \frac{1}{z}\int_z^1\!d\hat{z}\left(\frac{2E_{\hat{k}}(2\pi)^3d\hat{\sigma}}{d^3\vec{\hat{k}}}\right)\!\left(\xi^2\!\int \!d^2\vec{k}_{T}\, d(\xi,-\xi\vec{k}_T,\{P_h\})\right) + {\rm p.s.}\,. \label{e:fact_3}
\end{equation}
If we compare Eq.~(\ref{e:fact_3}) and Eq.~(\ref{e:fact_finalb}), using also Eqs.~(\ref{e:Jac}), (\ref{e:zxi}), we conclude the factorization formula in Ref.~\cite{Rogers:2024nhb} is just Eq.~(\ref{e:fact_finalb}) written using a $n$-hadron FF that is a number density in a set of variables  $(y_1,\dots,y_n, \vec{Y}_{1},\dots,\vec{Y}_{n})$ whose Jacobian when transforming from  $(\xi_1,\dots,\xi_n, \vec{P}_{1\perp},\dots,\vec{P}_{n\perp})$ is
\begin{equation}
    \left |\frac{\partial(\xi_1,\dots,\xi_n, \vec{P}_{1\perp},\dots,\vec{P}_{n\perp})}{\partial(y_1,\dots,y_n, \vec{Y}_{1},\dots,\vec{Y}_{n})} \right | = (16\pi^3)^{n-1}\,\frac{\xi_1\cdots\xi_n}{\xi}\,. \label{e:Jac_3}
\end{equation}
For general $n$, due to the ambiguous nature of $d(\xi,-\xi\vec{k}_T,\{P_h\})$ from Ref.~\cite{Rogers:2024nhb}, where it is not specified in which set of $3n$ variables $d(\xi,-\xi\vec{k}_T,\{P_h\})$ is a number density, we could not readily identify exactly what set of variables $(y_1,\dots,y_n, \vec{Y}_{1},\dots,\vec{Y}_{n})$ gives this Jacobian.  For the special case $n=2$, we were able to use the discussion in Ref.~\cite{Rogers:2024nhb} around Eqs.~(104)--(108), where (for the first time) explicit information can be found about what actually is meant by $d(\xi,\{P_h\})$ for the $n=2$ case, to recognize that 
\begin{equation}
    \left |\frac{\partial(\xi_1,\xi_2, \vec{P}_{1\perp},\vec{P}_{2\perp})}{\partial(\xi,\zeta, \xi\vec{k}_T,\vec{\widetilde{M}}_h)} \right | = 4\pi^3\xi(1-\zeta^2)=16\pi^3\, \frac{\xi_1\xi_2}{\xi}\,,
\end{equation}
where $\vec{\widetilde{M}}_h$ is a vector with magnitude $|\vec{\widetilde{M}}_h|=M_h/\sqrt{32\pi^3}$ and the same azimuthal angle $\phi_R$ as $\vec{R}_T$. That is, for $n=2$, $d(\xi,-\xi\vec{k}_T,\{P_h\})$ is a number density in $(\xi,\zeta,\xi\vec{k}_T,\vec{\widetilde{M}}_h)$.\footnote{Any set of variables with unit Jacobian relative to $(\xi,\zeta, \xi\vec{k}_T,\vec{\widetilde{M}}_h)$ could also be used.}  

Connected to this, Ref.~\cite{Pitonyak:2023gjx} was able to explicitly define DiFFs that are number densities in a given set of variables {\it before} performing parton model calculations.  Then it was shown that one obtains the expected result for the $e^+e^-\to (h_1h_2)\,X$ cross section differential in a given set of variables.  On the other hand, Ref.~\cite{Rogers:2024nhb} must define a ``reduced'' DiFF {\it in the middle of} a parton model calculation in such a way to ensure  the expected expression for the cross section is achieved.  This is a consequence of the fact that Ref.~\cite{Rogers:2024nhb} is not explicit in which set of $3n$ variables $d(\xi,-\xi\vec{k}_T,\{P_h\})$ is a number density.

To summarize, Eq.~(\ref{e:fact_finalb}) is a factorization formula for $e^+e^-\to (h_1\cdots h_n)\, X$ involving the fundamental $n$-hadron FF $D_1^{\{h_i\}_n}(\{\xi_i\}_n,\{\vec{P}_{i\perp}\}_n)$ from Ref.~\cite{Pitonyak:2023gjx}, which has an unambiguous interpretation as a number density in $(\xi_1,\dots,\xi_n, \vec{P}_{1\perp},\dots,\vec{P}_{n\perp})$.  In contrast, the factorization formula~\eqref{e:fact_3} above derived in Ref.~\cite{Rogers:2024nhb} involves a $n$-hadron FF $d(\xi,-\xi\vec{k}_T,\{P_h\})$ that is a number density in some other set of $3n$ variables which, for arbitray $n$, was not specified in Ref.~\cite{Rogers:2024nhb}, and we cannot readily identify them either. We have inferred, though, from comparing Eq.~(\ref{e:fact_finalb}) and Eq.~\eqref{e:fact_3}, that the Jacobian of the transformation from the variables $(\xi_1,\dots,\xi_n, \vec{P}_{1\perp},\dots,\vec{P}_{n\perp})$ to the variables used in $d(\xi,-\xi\vec{k}_T,\{P_h\})$ is given by Eq.~(\ref{e:Jac_3}).

%%%%%%%%%%%%%%%%%%%%%%%%%%%%%%%%%%%%%%%%%%
\section{Number Density Interpretation \label{s:num}}
%%%%%%%%%%%%%%%%%%%%%%%%%%%%%%%%%%%%%%%%%%
In Ref.~\cite{Pitonyak:2023gjx}, justification was given for $D_1^{\{h_i\}_n}(\{\xi_i\}_n,\{\vec{P}_{i\perp}\}_n)$ being a number density in $(\xi_1,\dots,\xi_n,$ $\vec{P}_{1\perp},\dots,\vec{P}_{n\perp})$ through a number sum rule, and for $n=2$ also a momentum sum rule and cross section calculations for $e^+e^-\to (h_1h_2)\,X$ (see Supplemental Material Secs.~S1 and S2 and the discussion around Eq.~(20) in the main text). In Ref.~\cite{Collins:2023cuo}, there was criticism that number sum rules for FFs do not hold in QCD, and Ref.~\cite{Rogers:2024nhb} argues therefore that factorization is the rigorous way in which one can develop the fundamental operator definition of $n$-hadron FFs.  In Appendix B of Ref.~\cite{Rogers:2024nhb}, an attempt is made to impose the number sum rule (B13) on a DiFF called $D_1(\xi,\zeta,\vec{k}_T,M_h^2)$ and a ``paradox'' is found with Eq.~(B9) calculated from a cross section.  Eventually  Ref.~\cite{Rogers:2024nhb} ``resolves'' this by stating Eq.~(B13) is not a valid relation.  However, 
the correct use of the number sum rule in Eq.~(6) of Ref.~\cite{Pitonyak:2023gjx} is for the integration measure in Eq.~(B13) of Ref.~\cite{Rogers:2024nhb} to be $d\xi\, d\zeta\, d^2(\xi\vec{k}_T) \,d^2 \vec{\widetilde{M}}_h$, since, as mentioned above, $D_1(\xi,\zeta,\vec{k}_T,M_h^2)$ is a number density in $(\xi,\zeta,\xi\vec{k}_T,\vec{\widetilde{M}}_h)$.  This then gives exactly the l.h.s.~of Eq.~(B9). 
Therefore, there is no contradiction between an operator definition of a $n$-hadron FF deduced from factorization and one based on a ``na\"{i}ve'' number sum rule.  The work here and in Ref.~\cite{Pitonyak:2023gjx} shows the two are fully compatible.

Even though Ref.~\cite{Rogers:2024nhb} relies on the factorization formula in Eq.~\eqref{e:fact_3} above to motivate its operator definition of a $n$-hadron FF,  its number density interpretation still must  be justified.  This is done by attempting to generalize the number density operator definition of a single-hadron FF discussed in Sec.~12.4 of Ref.~\cite{Collins:2011zzd}. If one looks at  Eq.~(12.35) of Ref.~\cite{Collins:2011zzd} (see also Eq.~(4.5) of Ref.~\cite{Collins:1981uw}), we can write\footnote{In Eq.~(12.35) of Ref.~\cite{Collins:2011zzd}, there is no $\sum_h$.  We have added it here for reasons that will be apparent when we generalize to $n$ hadrons.}
\begin{align}
    \sum_h D_1^h(\xi,\vec{P}_{\perp}) \langle \vec{k}_1|\vec{k}_2\rangle \equiv \langle \vec{k}_1|\frac{d\hat{N}}{d\xi d^2\vec{P}_{\perp}}|\vec{k}_2\rangle\,,
\end{align}
where $|\vec{k}_{1(2)}\rangle$ are quark states and $\hat{N}$ is the total number operator,
\begin{equation}
\hat{N}\equiv\sum_h\int\!\frac{d\xi\,d^2\vec{P}_{\perp}}{2\xi(2\pi)^3 }\hat{a}^\dagger_{h}\hat{a}_{h}\,,\label{e:num_op}
\end{equation}                           
with $\hat{a}^\dagger_{h}$ and $\hat{a}_{h}$ equal time creation and annihilation operators, respectively, for a hadron $h$ with momentum $\vec{P}$.  We highlight that the single-hadron FF $D_1^h(\xi,\vec{P}_{\perp})$ (summed over $h$) is defined as a number density in $(\xi,\vec{P}_\perp)$ by writing the total number operator as differential in those variables.  Starting from the operator definition of $D_1^{\{h_i\}_n}(\{\xi_i\}_n,\{\vec{P}_{i\perp}\}_n)$ in Eq.~\eqref{e:D1hn}, and using expressions for the quark field operator $\psi(x)$ in terms of (quark) lightcone creation and annihilation operators
as well as Eq.~(S30) of Ref.~\cite{Pitonyak:2023gjx}, we find
\begin{equation}
    \sum_{h_1}\cdots \sum_{h_n} D_1^{\{h_i\}_n}(\{\xi_i\}_n,\{\vec{P}_{i\perp}\}_n) \langle \vec{k}_1|\vec{k}_2\rangle = \langle \vec{k}_1|\frac{d\left( \prod_{\scriptscriptstyle k=0}^{\scriptscriptstyle n-1}(\hat{N}-k)\right)}{d\xi_1\cdots d\xi_n d^2\vec{P}_{1\perp}\cdots d^2\vec{P}_{n\perp}}|\vec{k}_2\rangle\,.\label{e:dN}
\end{equation}
We recognize on the r.h.s.~the operator for the total number of $n$-hadrons differential in $(\xi_1,\dots,\xi_n, \vec{P}_{1\perp},\dots,\vec{P}_{n\perp})$, which is exactly what one wants for the definition of an $n$-hadron FF that is a number density in those variables.  This is an additional result to what was already derived in Ref.~\cite{Pitonyak:2023gjx} to justify the number density interpretation of $D_1^{\{h_i\}_n}(\{\xi_i\}_n,\{\vec{P}_{i\perp}\}_n)$. If we follow the same steps for the operator definition of $d(\xi,-\xi\vec{k}_T,\{P_h\})$ in Eq.~(\ref{e:D1hn_79}), we instead find the r.h.s.~of Eq.~(\ref{e:dN}) is modified by the Jacobian factor in Eq.~(\ref{e:Jac_3}) (see also the second line of Eq.~(77) in Ref.~\cite{Rogers:2024nhb}),
which for $n\geq 2$ does not give the standard expression we expect for a number density.   We therefore disagree that Eq.~(77) of Ref.~\cite{Rogers:2024nhb} is how to generalize the fundamental single-hadron FF to the $n$-hadron case. 

%%%%%%%%%%%%%%%%%%%%%%%%%%%%%%%%%%%%%%%%%%
\section{Dihadron Fragmentation Evolution Equations \label{s:evo}}
%%%%%%%%%%%%%%%%%%%%%%%%%%%%%%%%%%%%%%%%%%
In the ``Evolution of Extended DiFFs'' section of Ref.~\cite{Pitonyak:2023gjx} as well as Supplemental Material Sec.~S4, it was explained how to derive the evolution equations for the DiFF definitions there, and it was found that the splitting functions are the same as for the single-hadron case.  The only potential change is in the measure of the convolution integral, depending on which DiFF is under consideration. To highlight this, here are the evolution equations (to $\mathcal{O}(\alpha_s)$) derived in Ref.~\cite{Pitonyak:2023gjx} for $D_1^{h_1h_2}(\xi,M_h)$ and $D_1^{h_1h_2}(\xi_1,\xi_2,\vec{R}_T)$:
\begin{subequations}
\begin{align}
     \frac{\partial D_1^{h_1h_2/i}(\xi,M_h;\mu)}{\partial \ln\mu^2} &=\sum_{i'}\!\int_{\xi}^1\! \frac{d\hat{z}}{\hat{z}}D_1^{h_1h_2/i'}\!\!\left(\!\frac{\xi}{\hat{z}},M_h;\mu\!\right)\!P_{i\to i'}(\hat{z})\,, \label{e:D1zMhevo}\\[0.3cm]
     \frac{\partial D_1^{h_1h_2/i}(\xi_1,\xi_2,\vec{R}_T;\mu)}{\partial \ln\mu^2}&= \sum_{i'}\!\int_{\xi}^1\! \frac{d\hat{z}}{\hat{z}^2}D_1^{h_1h_2/i'}\!\!\left(\!\frac{\xi_1}{\hat{z}},\frac{\xi_2}{\hat{z}},\vec{R}_T;\mu\!\right)\!P_{i\to i'}(\hat{z})\,, \label{e:D1z1z2evo}
\end{align}
\end{subequations}
where $P_{i\to i'}(\hat{z})$ are the standard unpolarized time-like splitting functions~\cite{Altarelli:1977zs}. We note that the integration measure in (\ref{e:D1zMhevo}) is $d\hat{z}/\hat{z}$ while in (\ref{e:D1z1z2evo}) it is $d\hat{z}/\hat{z}^2$.  The latter matches the first line in Eq.~(30) of Ref.~\cite{deFlorian:2003cg} and the first and fourth lines of Eq.~(23) of Ref.~\cite{Majumder:2004br} as well as Eq.~(24) of Ref.~\cite{Ceccopieri:2007ip}.   We also mention that (\ref{e:D1z1z2evo}) follows from (\ref{e:D1zMhevo}) (or vice versa) using Eqs.~(\ref{e:D1z1z2RT}), (\ref{e:D1zMh}), where the fact that $\xi_1,\xi_2$ in the definition of $D_1^{h_1h_2}(\xi_1,\xi_2,\vec{R}_T)$ are fractions of the parton's momentum carried by the hadrons is crucial in this regard.  (They were instead identified as external kinematic variables in Eq.~(130) of Ref.~\cite{Rogers:2024nhb}.)  Appendix~C of Ref.~\cite{Rogers:2024nhb} essentially recognized that a simple change in integration measure of the evolution equations occurs.  However, it was interpreted as modifying the splitting function because an incorrect factorization formula was being used with the DiFFs from Ref.~\cite{Pitonyak:2023gjx}.  The integration measures in a NLO calculation of our DiFF correlators would be the same as our evolution equations (\ref{e:D1zMhevo}), (\ref{e:D1z1z2evo}) and are consistent with the integration measure in the factorization formula, Eq.~(\ref{e:fact_finalb}), which is needed in order to be able to cancel the collinear singularities in the cross section that arise at NLO.  Namely, we find
\begin{subequations}
\begin{align}
    \frac{d\sigma}{dz\,dM_h}&=\int_z^1 \!\frac{d\hat{z}}{\hat{z}}\frac{d\hat{\sigma}}{d\hat{z}} D_1^{h_1h_2}\!\left(\frac{z}{\hat{z}},M_h\right), \label{e:dsig_gena}\\[0.3cm]
    \frac{d\sigma}{dz_1dz_2d^2\vec{R}_T}&=\int_z^1 \!\frac{d\hat{z}} {\hat{z}^2}\frac{d\hat{\sigma}}{d\hat{z}}D_1^{h_1h_2}\!\left(\frac{z_1}{\hat{z}},\frac{z_2}{\hat{z}},\vec{R}_T\right). \label{e:dsig_genb}
\end{align}
\end{subequations}
The change in integration measure compared to the single-hadron case was also discussed previously for $d\sigma/dz_1dz_2$ in Ref.~\cite{Majumder:2004br} (see after Eq.~(23)).  
Note that if one demands that the convolution integral in Eq.~\eqref{e:dsig_genb} be $\int_z^1\!\frac{d\hat{z}}{\hat{z}}\,\frac{d\hat{\sigma}}{d\hat{z}}\cdots$, as it is for the single-hadron case, then one is forced to define a DiFF that is not a number density in $(\xi_1,\xi_2)=(\frac{z_1}{\hat{z}},\frac{z_2}{\hat{z}})$ 
(or to modify the hard factor by including $1/\hat{z}$ in it).  We also remark that one can derive Eq.~\eqref{e:dsig_gena} starting from Eq.~\eqref{e:dsig_genb} by rewriting the l.h.s.~of \eqref{e:dsig_genb} using the Jacobian for the variable transformation from $(z,\zeta,\vec{M}_h)$ to $(z_1,z_2,\vec{R}_T)$ and then solving for $d\sigma/dz\,dM_h$ ($\vec{M}_h$ is a vector with magnitude $M_h$ and azimuthal angle $\phi_{R_T}$), making use of the definition of $D_1^{h_1h_2}(\xi,M_h)$ in Eq.~\eqref{e:D1zMh}.

%%%%%%%%%%%%%%%%%%%%%%%%%%%%%%%%%%%%%%%%%%
\section{Previous Definitions of $\boldsymbol{D_1^{h_1h_2}(\xi,M_h)}$ and Unpolarized \newline Cross Section Formulas in the Literature \label{s:prevdef}}
%%%%%%%%%%%%%%%%%%%%%%%%%%%%%%%%%%%%%%%%%%
A DiFF of particular interest for phenomenology is $D_1^{h_1h_2}(\xi,M_h)$.\footnote{We mention that in the literature the variable $\xi=(1+\zeta)/2$ has sometimes been used, which is different from the momentum fraction variable $\xi$ that we have been using here.} An operator definition of this function was provided in Eq.~(17) of Ref.~\cite{Pitonyak:2023gjx}, which is also written explicitly in terms of $\Delta^{h_1h_2}$ in Eq.~(\ref{e:D1zMh}) above.  Using Eqs.~(108), (105), and (81) from Ref.~\cite{Rogers:2024nhb}, one readily finds that $d_{\rm red, 1}(\xi,M_h)$ defined there exactly matches Eq.~(\ref{e:D1zMh}).  In Ref.~\cite{Rogers:2024nhb}, it was implied that this definition appeared previously in the literature, e.g., in Ref.~\cite{Bacchetta:2003vn}.  In the context of $D_1^{h_1h_2}(\xi,M_h)$, Ref.~\cite{Bacchetta:2003vn} cites Ref.~\cite{Bacchetta:2002ux}.  By using Eqs.~(32), (29), (16), (7) of Ref.~\cite{Bacchetta:2002ux}, we attempted to write down the operator definition of $D_1^{h_1h_2}(\xi,M_h)$ in terms of $\Delta^{h_1h_2}$ in Eq.~(\ref{e:DiFF_corr}) and found
\begin{equation}
    D_1^{h_1h_2}(\xi,M_h)|_{\text{Ref.\,\cite{Bacchetta:2002ux}}}=\frac{\pi\xi}{16}\int\!d\zeta\int\! d^2\vec{k}_T\,\Delta^{h_1h_2}\,,\label{e:D1_02}
\end{equation}
which does not match Eq.~(\ref{e:D1zMh}). Similarly, we could not find any paper from Refs.~\cite{Radici:2001na,Boer:2003ya,Gliske:2014wba,Matevosyan:2018icf} that, after attempting to write the version of $D_1^{h_1h_2}(\xi,M_h)$ there in terms of the correlator $\Delta^{h_1h_2}$, gave us Eq.~(\ref{e:D1zMh}). We note that the l.h.s.~of Eq.~(\ref{e:D1_02}) is what was denoted in Eq.~(32) of Ref.~\cite{Bacchetta:2002ux} as $D_{1,OO}(\xi,M_h^2)$ ($\xi$ here being the same as $z$ in Ref.~\cite{Bacchetta:2002ux}).  Given the statements after Eq.~(110) of Ref.~\cite{Rogers:2024nhb}, where the DiFF $d_{\rm red, 1}(\xi,M_h)$ enters, we understood $D_{1,OO}(\xi,M_h^2)$ as attempting to define a number density in $(\xi,M_h)$.  If instead an attempt was being made to define a number density in $(\xi,M_h^2)$, this is still incorrect, as one can see by multiplying r.h.s.~of Eq.~(\ref{e:D1zMh}) by the Jacobian $1/(2M_h)$ and comparing it to the r.h.s.~of Eq.~(\ref{e:D1_02}).  In addition, we mention that Eq.~(B10) of Ref.~\cite{Rogers:2024nhb} is not how Refs.~\cite{Bacchetta:2002ux,Bacchetta:2003vn} and related references go from $D_1(\xi,\zeta,M_h)$ to $D_1(\xi,M_h)$:~it is instead done through a Legendre expansion in $\zeta$ and then $D_1(\xi,M_h)\equiv \frac{1}{2}\!\int\! d\zeta \,D_1(\xi,\zeta,M_h)$. 

Moreover, there seem to be inconsistencies in how $D_1^{h_1h_2}(\xi,M_h)$ enters previous unpolarized cross section formulas in the literature.  In Ref.~\cite{Rogers:2024nhb}, it was emphasized that Eq.~(9) of Ref.~\cite{Courtoy:2012ry}\footnote{We note that Eq.~(9) in Ref.~\cite{Courtoy:2012ry} was not calculated in the same way as in Ref.~\cite{Rogers:2024nhb}.  In Ref.~\cite{Courtoy:2012ry}, it was stated the starting point is Eq.~(30) of Ref.~\cite{Boer:2003ya}, which, attempting to follow from there, we worked out to give $\frac{d\sigma}{dz\,dM_h}=\frac{4\pi N_c\alpha_{em}^2}{3Q^2} \sum_q e_q^2\,[2M_h z^2 D_1^{h_1h_2/q} (z,M_h^2)|_\text{Ref.\,\cite{Boer:2003ya}}]$.  We also tried starting from Eq.~(45) of Ref.~\cite{Matevosyan:2018icf} and similarly found a different expression than Eq.~(9) of Ref.~\cite{Courtoy:2012ry}, with  the factor in brackets now being $[M_h/(2\pi^2) D_{1,OO}^{h_1h_2/q} (z,M_h^2)|_\text{Ref.\,\cite{Matevosyan:2018icf}}]$.  We further mention that the cross section formula in Eq.~(9) of Ref.~\cite{Courtoy:2012ry} is differential in $z$, $M_h$, {\it and} $Q^2$, which gives $D_1(z,M_h)$ there different dimensions than $d_{\rm red, 1}(z,M_h)$ of Ref.~\cite{Rogers:2024nhb}.}, used to extract $D_1^{h_1h_2}(\xi,M_h)$ from $d\sigma/dz\,dM_h$ data for $e^+e^-\to (h_1h_2)\,X$, agrees with Eq.~(110) of Ref.~\cite{Rogers:2024nhb} (which is the same as Eq.~(20) of Ref.~\cite{Pitonyak:2023gjx}) and is the expected parton model result if $D_1^{h_1h_2}(\xi,M_h)$ has been correctly defined as a number density in $(\xi,M_h)$.  Another process where DiFFs enter is  semi-inclusive deep-inelastic scattering (SIDIS), $e\,N\to e\,(h_1h_2)\,X$\footnote{We refer the reader to, e.g., Refs.~\cite{Bacchetta:2003vn,Bacchetta:2006tn} for the standard setup and kinematic variable definitions in SIDIS.}, which was not studied in Refs.~\cite{Rogers:2024nhb,Pitonyak:2023gjx}.  Using the definition of $D_1^{h_1h_2}(\xi,M_h)$ in Eq.~(\ref{e:D1zMh}), we calculated the leading-order unpolarized SIDIS cross section and found
\begin{equation}
    \frac{d\sigma}{dx\,dy\,dz\,dM_h} = \frac{4\pi\alpha_{em}^2}{yQ^2}(1-y+y^2/2)\sum_q e_q^2\,f_1^{q/N}(x)\,D_1^{h_1h_2/q}(z,M_h)|_\text{Ref.\,\cite{Pitonyak:2023gjx}}\,. \label{e:sidis}
\end{equation}
This is exactly what one expects if $D_1^{h_1h_2}(\xi,M_h)$ is a number density in $(\xi,M_h)$ (cf.~Eq.~(12.91) of Ref.~\cite{Collins:2011zzd}).  On the other hand, starting from Eq.~(2.5) of Ref.~\cite{Radici:2015mwa} (see also Eq.~(2.6) of Ref.~\cite{Bacchetta:2012ty}), the result instead reads
\begin{equation}
    \frac{d\sigma}{dx\,dy\,dz\,dM_h} = \frac{4\pi\alpha_{em}^2}{yQ^2}(1-y+y^2/2)\sum_q e_q^2\,f_1^{q/N}(x)\left[4\pi M_h\,D_1^{h_1h_2/q}(z,M_h)|_\text{Ref.\,\cite{Radici:2015mwa}}\right], \label{e:sidis_15}
\end{equation}
which means $D_1^{h_1h_2}(\xi,M_h)$ used in that formula is not a number density in $(\xi,M_h)$.  (Note that we identified $D_1^{h_1h_2}(z,M_h)|_\text{Ref.\,\cite{Radici:2015mwa}}\equiv\frac{1}{2}\!\int\!d(\cos\theta)\,D_1^{h_1h_2}(z,\cos\theta, M_h)|_\text{Ref.\,\cite{Radici:2015mwa}}$, which also aligns with the discussion surrounding Eq.~(2.12) of Ref.~\cite{Bacchetta:2012ty}.)  Therefore, Eq.~(2.5) of Ref.~\cite{Radici:2015mwa} seems to be inconsistent with Eq.~(9) of Ref.~\cite{Courtoy:2012ry} in terms of the DiFF that is used.

%%%%%%%%%%%%%%%%%%%%%%%%%%%%%%%%%%%%%%%%%%
\section{Summary and Conclusions \label{s:concl}}
%%%%%%%%%%%%%%%%%%%%%%%%%%%%%%%%%%%%%%%%%%
We have reconfirmed that the DiFF (and $n$-hadron FF) definitions in Ref.~\cite{Pitonyak:2023gjx} have a natural number density interpretation and are consistent with collinear factorization using the standard hard factors and evolution kernels, with Eqs.~(\ref{e:fact_finalb}), (\ref{e:dN}), (\ref{e:sidis}) here adding to and supporting the results from Ref.~\cite{Pitonyak:2023gjx}.   We have also corrected erroneous claims in Ref.~\cite{Rogers:2024nhb} about our work in Ref.~\cite{Pitonyak:2023gjx} as well as pointed out some inconsistencies in previous DiFF definitions and unpolarized cross section formulas in the literature.  
We emphasize that the approach in Ref.~\cite{Pitonyak:2023gjx} of using sum rules and parton model calculations to support the definition of DiFFs (or $n$-hadron FFs) as number densities remains valid.  There have been no new insights offered in that regard by the analysis of (leading order) factorization in Ref.~\cite{Rogers:2024nhb} (and the introduction of an ambiguous $n$-hadron FF $d(\xi,-\xi\vec{k}_T,\{P_h\})$). With respect to factorization, establishing a proof for $e^+e^-\to (h_1\cdots h_n)\, X$ to all orders, and valid for all $M_h$, that includes contributions involving FFs for less than $n$ hadrons still remains an open problem.
%%%%%%%%%%%%%%%%%%%%%%%%%%%%%%%%%%%%%%%%%%
\begin{acknowledgments}
This work was supported by the National Science Foundation under Grants No.~PHY-2110472, No.~PHY-2412792 (A.M.), No.~PHY-2308567 (D.P.), and No.~PHY-2310031, No.~PHY-2335114 (A.P.), and the U.S. Department of Energy contract No.~DE-AC05-06OR23177, under which Jefferson Science Associates, LLC operates Jefferson Lab (A.P.~and N.S.). 
The work of N.S. and C.C.~was also supported by the DOE, Office of Science, Office of Nuclear Physics in the Early Career Program.
\end{acknowledgments}
%%%%%%%%%%%%%%%%%%%%%%%%%%%%%%%%%%%%%%%%%%



\begin{thebibliography}{22}%
\makeatletter
\providecommand \@ifxundefined [1]{%
 \@ifx{#1\undefined}
}%
\providecommand \@ifnum [1]{%
 \ifnum #1\expandafter \@firstoftwo
 \else \expandafter \@secondoftwo
 \fi
}%
\providecommand \@ifx [1]{%
 \ifx #1\expandafter \@firstoftwo
 \else \expandafter \@secondoftwo
 \fi
}%
\providecommand \natexlab [1]{#1}%
\providecommand \enquote  [1]{``#1''}%
\providecommand \bibnamefont  [1]{#1}%
\providecommand \bibfnamefont [1]{#1}%
\providecommand \citenamefont [1]{#1}%
\providecommand \href@noop [0]{\@secondoftwo}%
\providecommand \href [0]{\begingroup \@sanitize@url \@href}%
\providecommand \@href[1]{\@@startlink{#1}\@@href}%
\providecommand \@@href[1]{\endgroup#1\@@endlink}%
\providecommand \@sanitize@url [0]{\catcode `\\12\catcode `\$12\catcode
  `\&12\catcode `\#12\catcode `\^12\catcode `\_12\catcode `\%12\relax}%
\providecommand \@@startlink[1]{}%
\providecommand \@@endlink[0]{}%
\providecommand \url  [0]{\begingroup\@sanitize@url \@url }%
\providecommand \@url [1]{\endgroup\@href {#1}{\urlprefix }}%
\providecommand \urlprefix  [0]{URL }%
\providecommand \Eprint [0]{\href }%
\providecommand \doibase [0]{http://dx.doi.org/}%
\providecommand \selectlanguage [0]{\@gobble}%
\providecommand \bibinfo  [0]{\@secondoftwo}%
\providecommand \bibfield  [0]{\@secondoftwo}%
\providecommand \translation [1]{[#1]}%
\providecommand \BibitemOpen [0]{}%
\providecommand \bibitemStop [0]{}%
\providecommand \bibitemNoStop [0]{.\EOS\space}%
\providecommand \EOS [0]{\spacefactor3000\relax}%
\providecommand \BibitemShut  [1]{\csname bibitem#1\endcsname}%
\let\auto@bib@innerbib\@empty
%</preamble>
\bibitem [{\citenamefont {Rogers}\ \emph {et~al.}(2025)\citenamefont {Rogers},
  \citenamefont {Radici}, \citenamefont {Courtoy},\ and\ \citenamefont
  {Rainaldi}}]{Rogers:2024nhb}%
  \BibitemOpen
  \bibfield  {author} {\bibinfo {author} {\bibfnamefont {T.~C.}\ \bibnamefont
  {Rogers}}, \bibinfo {author} {\bibfnamefont {M.}~\bibnamefont {Radici}},
  \bibinfo {author} {\bibfnamefont {A.}~\bibnamefont {Courtoy}}, \ and\
  \bibinfo {author} {\bibfnamefont {T.}~\bibnamefont {Rainaldi}},\ }\href
  {\doibase 10.1103/PhysRevD.111.056001} {\bibfield  {journal} {\bibinfo
  {journal} {Phys. Rev. D}\ }\textbf {\bibinfo {volume} {111}},\ \bibinfo
  {pages} {056001} (\bibinfo {year} {2025})},\ \Eprint
  {http://arxiv.org/abs/2412.12282} {arXiv:2412.12282 [hep-ph]} \BibitemShut
  {NoStop}%
\bibitem [{\citenamefont {Pitonyak}\ \emph {et~al.}(2024)\citenamefont
  {Pitonyak}, \citenamefont {Cocuzza}, \citenamefont {Metz}, \citenamefont
  {Prokudin},\ and\ \citenamefont {Sato}}]{Pitonyak:2023gjx}%
  \BibitemOpen
  \bibfield  {author} {\bibinfo {author} {\bibfnamefont {D.}~\bibnamefont
  {Pitonyak}}, \bibinfo {author} {\bibfnamefont {C.}~\bibnamefont {Cocuzza}},
  \bibinfo {author} {\bibfnamefont {A.}~\bibnamefont {Metz}}, \bibinfo {author}
  {\bibfnamefont {A.}~\bibnamefont {Prokudin}}, \ and\ \bibinfo {author}
  {\bibfnamefont {N.}~\bibnamefont {Sato}},\ }\href {\doibase
  10.1103/PhysRevLett.132.011902} {\bibfield  {journal} {\bibinfo  {journal}
  {Phys. Rev. Lett.}\ }\textbf {\bibinfo {volume} {132}},\ \bibinfo {pages}
  {011902} (\bibinfo {year} {2024})},\ \Eprint
  {http://arxiv.org/abs/2305.11995} {arXiv:2305.11995 [hep-ph]} \BibitemShut
  {NoStop}%
\bibitem [{\citenamefont {Collins}\ and\ \citenamefont
  {Soper}(1982)}]{Collins:1981uw}%
  \BibitemOpen
  \bibfield  {author} {\bibinfo {author} {\bibfnamefont {J.~C.}\ \bibnamefont
  {Collins}}\ and\ \bibinfo {author} {\bibfnamefont {D.~E.}\ \bibnamefont
  {Soper}},\ }\href@noop {} {\bibfield  {journal} {\bibinfo  {journal} {Nucl.
  Phys.}\ }\textbf {\bibinfo {volume} {B194}},\ \bibinfo {pages} {445}
  (\bibinfo {year} {1982})}\BibitemShut {NoStop}%
%%CITATION = NUPHA,B194,445;%%
\bibitem [{\citenamefont {Metz}\ and\ \citenamefont
  {Vossen}(2016)}]{Metz:2016swz}%
  \BibitemOpen
  \bibfield  {author} {\bibinfo {author} {\bibfnamefont {A.}~\bibnamefont
  {Metz}}\ and\ \bibinfo {author} {\bibfnamefont {A.}~\bibnamefont {Vossen}},\
  }\href {\doibase 10.1016/j.ppnp.2016.08.003} {\bibfield  {journal} {\bibinfo
  {journal} {Prog. Part. Nucl. Phys.}\ }\textbf {\bibinfo {volume} {91}},\
  \bibinfo {pages} {136} (\bibinfo {year} {2016})},\ \Eprint
  {http://arxiv.org/abs/1607.02521} {arXiv:1607.02521 [hep-ex]} \BibitemShut
  {NoStop}%
%%CITATION = ARXIV:1607.02521;%%
\bibitem [{\citenamefont {Collins}(2011)}]{Collins:2011zzd}%
  \BibitemOpen
  \bibfield  {author} {\bibinfo {author} {\bibfnamefont {J.}~\bibnamefont
  {Collins}},\ }\href@noop {} {\bibfield  {journal} {\bibinfo  {journal} {Camb.
  Monogr. Part. Phys. Nucl. Phys. Cosmol.}\ }\textbf {\bibinfo {volume} {32}},\
  \bibinfo {pages} {1} (\bibinfo {year} {2011})}\BibitemShut {NoStop}%
%%CITATION = CMPCE,32,1;%%
\bibitem [{\citenamefont {Bianconi}\ \emph {et~al.}(2000)\citenamefont
  {Bianconi}, \citenamefont {Boffi}, \citenamefont {Jakob},\ and\ \citenamefont
  {Radici}}]{Bianconi:1999cd}%
  \BibitemOpen
  \bibfield  {author} {\bibinfo {author} {\bibfnamefont {A.}~\bibnamefont
  {Bianconi}}, \bibinfo {author} {\bibfnamefont {S.}~\bibnamefont {Boffi}},
  \bibinfo {author} {\bibfnamefont {R.}~\bibnamefont {Jakob}}, \ and\ \bibinfo
  {author} {\bibfnamefont {M.}~\bibnamefont {Radici}},\ }\href {\doibase
  10.1103/PhysRevD.62.034008} {\bibfield  {journal} {\bibinfo  {journal} {Phys.
  Rev.}\ }\textbf {\bibinfo {volume} {D62}},\ \bibinfo {pages} {034008}
  (\bibinfo {year} {2000})},\ \Eprint {http://arxiv.org/abs/hep-ph/9907475}
  {arXiv:hep-ph/9907475 [hep-ph]} \BibitemShut {NoStop}%
%%CITATION = HEP-PH/9907475;%%
\bibitem [{\citenamefont {Collins}\ and\ \citenamefont
  {Rogers}(2024)}]{Collins:2023cuo}%
  \BibitemOpen
  \bibfield  {author} {\bibinfo {author} {\bibfnamefont {J.}~\bibnamefont
  {Collins}}\ and\ \bibinfo {author} {\bibfnamefont {T.~C.}\ \bibnamefont
  {Rogers}},\ }\href {\doibase 10.1103/PhysRevD.109.016006} {\bibfield
  {journal} {\bibinfo  {journal} {Phys. Rev. D}\ }\textbf {\bibinfo {volume}
  {109}},\ \bibinfo {pages} {016006} (\bibinfo {year} {2024})},\ \Eprint
  {http://arxiv.org/abs/2309.03346} {arXiv:2309.03346 [hep-ph]} \BibitemShut
  {NoStop}%
\bibitem [{\citenamefont {de~Florian}\ and\ \citenamefont
  {Vanni}(2004)}]{deFlorian:2003cg}%
  \BibitemOpen
  \bibfield  {author} {\bibinfo {author} {\bibfnamefont {D.}~\bibnamefont
  {de~Florian}}\ and\ \bibinfo {author} {\bibfnamefont {L.}~\bibnamefont
  {Vanni}},\ }\href {\doibase 10.1016/j.physletb.2003.10.047} {\bibfield
  {journal} {\bibinfo  {journal} {Phys. Lett. B}\ }\textbf {\bibinfo {volume}
  {578}},\ \bibinfo {pages} {139} (\bibinfo {year} {2004})},\ \Eprint
  {http://arxiv.org/abs/hep-ph/0310196} {arXiv:hep-ph/0310196} \BibitemShut
  {NoStop}%
\bibitem [{\citenamefont {Majumder}\ and\ \citenamefont
  {Wang}(2005)}]{Majumder:2004br}%
  \BibitemOpen
  \bibfield  {author} {\bibinfo {author} {\bibfnamefont {A.}~\bibnamefont
  {Majumder}}\ and\ \bibinfo {author} {\bibfnamefont {X.-N.}\ \bibnamefont
  {Wang}},\ }\href {\doibase 10.1103/PhysRevD.72.034007} {\bibfield  {journal}
  {\bibinfo  {journal} {Phys. Rev. D}\ }\textbf {\bibinfo {volume} {72}},\
  \bibinfo {pages} {034007} (\bibinfo {year} {2005})},\ \Eprint
  {http://arxiv.org/abs/hep-ph/0411174} {arXiv:hep-ph/0411174} \BibitemShut
  {NoStop}%
\bibitem [{\citenamefont {Majumder}\ and\ \citenamefont
  {Wang}(2004)}]{Majumder:2004wh}%
  \BibitemOpen
  \bibfield  {author} {\bibinfo {author} {\bibfnamefont {A.}~\bibnamefont
  {Majumder}}\ and\ \bibinfo {author} {\bibfnamefont {X.-N.}\ \bibnamefont
  {Wang}},\ }\href {\doibase 10.1103/PhysRevD.70.014007} {\bibfield  {journal}
  {\bibinfo  {journal} {Phys. Rev. D}\ }\textbf {\bibinfo {volume} {70}},\
  \bibinfo {pages} {014007} (\bibinfo {year} {2004})},\ \Eprint
  {http://arxiv.org/abs/hep-ph/0402245} {arXiv:hep-ph/0402245} \BibitemShut
  {NoStop}%
\bibitem [{\citenamefont {Altarelli}\ and\ \citenamefont
  {Parisi}(1977)}]{Altarelli:1977zs}%
  \BibitemOpen
  \bibfield  {author} {\bibinfo {author} {\bibfnamefont {G.}~\bibnamefont
  {Altarelli}}\ and\ \bibinfo {author} {\bibfnamefont {G.}~\bibnamefont
  {Parisi}},\ }\href@noop {} {\bibfield  {journal} {\bibinfo  {journal} {Nucl.
  Phys.}\ }\textbf {\bibinfo {volume} {B126}},\ \bibinfo {pages} {298}
  (\bibinfo {year} {1977})}\BibitemShut {NoStop}%
%%CITATION = NUPHA,B126,298;%%
\bibitem [{\citenamefont {Ceccopieri}\ \emph {et~al.}(2007)\citenamefont
  {Ceccopieri}, \citenamefont {Radici},\ and\ \citenamefont
  {Bacchetta}}]{Ceccopieri:2007ip}%
  \BibitemOpen
  \bibfield  {author} {\bibinfo {author} {\bibfnamefont {F.~A.}\ \bibnamefont
  {Ceccopieri}}, \bibinfo {author} {\bibfnamefont {M.}~\bibnamefont {Radici}},
  \ and\ \bibinfo {author} {\bibfnamefont {A.}~\bibnamefont {Bacchetta}},\
  }\href {\doibase 10.1016/j.physletb.2007.04.065} {\bibfield  {journal}
  {\bibinfo  {journal} {Phys. Lett. B}\ }\textbf {\bibinfo {volume} {650}},\
  \bibinfo {pages} {81} (\bibinfo {year} {2007})},\ \Eprint
  {http://arxiv.org/abs/hep-ph/0703265} {arXiv:hep-ph/0703265} \BibitemShut
  {NoStop}%
\bibitem [{\citenamefont {Bacchetta}\ and\ \citenamefont
  {Radici}(2004)}]{Bacchetta:2003vn}%
  \BibitemOpen
  \bibfield  {author} {\bibinfo {author} {\bibfnamefont {A.}~\bibnamefont
  {Bacchetta}}\ and\ \bibinfo {author} {\bibfnamefont {M.}~\bibnamefont
  {Radici}},\ }\href {\doibase 10.1103/PhysRevD.69.074026} {\bibfield
  {journal} {\bibinfo  {journal} {Phys. Rev. D}\ }\textbf {\bibinfo {volume}
  {69}},\ \bibinfo {pages} {074026} (\bibinfo {year} {2004})},\ \Eprint
  {http://arxiv.org/abs/hep-ph/0311173} {arXiv:hep-ph/0311173} \BibitemShut
  {NoStop}%
\bibitem [{\citenamefont {Bacchetta}\ and\ \citenamefont
  {Radici}(2003)}]{Bacchetta:2002ux}%
  \BibitemOpen
  \bibfield  {author} {\bibinfo {author} {\bibfnamefont {A.}~\bibnamefont
  {Bacchetta}}\ and\ \bibinfo {author} {\bibfnamefont {M.}~\bibnamefont
  {Radici}},\ }\href {\doibase 10.1103/PhysRevD.67.094002} {\bibfield
  {journal} {\bibinfo  {journal} {Phys. Rev. D}\ }\textbf {\bibinfo {volume}
  {67}},\ \bibinfo {pages} {094002} (\bibinfo {year} {2003})},\ \Eprint
  {http://arxiv.org/abs/hep-ph/0212300} {arXiv:hep-ph/0212300} \BibitemShut
  {NoStop}%
\bibitem [{\citenamefont {Radici}\ \emph {et~al.}(2002)\citenamefont {Radici},
  \citenamefont {Jakob},\ and\ \citenamefont {Bianconi}}]{Radici:2001na}%
  \BibitemOpen
  \bibfield  {author} {\bibinfo {author} {\bibfnamefont {M.}~\bibnamefont
  {Radici}}, \bibinfo {author} {\bibfnamefont {R.}~\bibnamefont {Jakob}}, \
  and\ \bibinfo {author} {\bibfnamefont {A.}~\bibnamefont {Bianconi}},\ }\href
  {\doibase 10.1103/PhysRevD.65.074031} {\bibfield  {journal} {\bibinfo
  {journal} {Phys. Rev.}\ }\textbf {\bibinfo {volume} {D65}},\ \bibinfo {pages}
  {074031} (\bibinfo {year} {2002})},\ \Eprint
  {http://arxiv.org/abs/hep-ph/0110252} {arXiv:hep-ph/0110252 [hep-ph]}
  \BibitemShut {NoStop}%
%%CITATION = HEP-PH/0110252;%%
\bibitem [{\citenamefont {Boer}\ \emph {et~al.}(2003)\citenamefont {Boer},
  \citenamefont {Jakob},\ and\ \citenamefont {Radici}}]{Boer:2003ya}%
  \BibitemOpen
  \bibfield  {author} {\bibinfo {author} {\bibfnamefont {D.}~\bibnamefont
  {Boer}}, \bibinfo {author} {\bibfnamefont {R.}~\bibnamefont {Jakob}}, \ and\
  \bibinfo {author} {\bibfnamefont {M.}~\bibnamefont {Radici}},\ }\href
  {\doibase 10.1103/PhysRevD.67.094003} {\bibfield  {journal} {\bibinfo
  {journal} {Phys. Rev.}\ }\textbf {\bibinfo {volume} {D67}},\ \bibinfo {pages}
  {094003} (\bibinfo {year} {2003})},\ \Eprint
  {http://arxiv.org/abs/hep-ph/0302232} {arXiv:hep-ph/0302232 [hep-ph]}
  \BibitemShut {NoStop}%
%%CITATION = HEP-PH/0302232;%%
\bibitem [{\citenamefont {Gliske}\ \emph {et~al.}(2014)\citenamefont {Gliske},
  \citenamefont {Bacchetta},\ and\ \citenamefont {Radici}}]{Gliske:2014wba}%
  \BibitemOpen
  \bibfield  {author} {\bibinfo {author} {\bibfnamefont {S.}~\bibnamefont
  {Gliske}}, \bibinfo {author} {\bibfnamefont {A.}~\bibnamefont {Bacchetta}}, \
  and\ \bibinfo {author} {\bibfnamefont {M.}~\bibnamefont {Radici}},\ }\href
  {\doibase 10.1103/PhysRevD.90.114027} {\bibfield  {journal} {\bibinfo
  {journal} {Phys. Rev. D}\ }\textbf {\bibinfo {volume} {90}},\ \bibinfo
  {pages} {114027} (\bibinfo {year} {2014})},\ \bibinfo {note} {[Erratum:
  Phys.Rev.D 91, 019902 (2015)]},\ \Eprint {http://arxiv.org/abs/1408.5721}
  {arXiv:1408.5721 [hep-ph]} \BibitemShut {NoStop}%
\bibitem [{\citenamefont {Matevosyan}\ \emph {et~al.}(2018)\citenamefont
  {Matevosyan}, \citenamefont {Bacchetta}, \citenamefont {Boer}, \citenamefont
  {Courtoy}, \citenamefont {Kotzinian}, \citenamefont {Radici},\ and\
  \citenamefont {Thomas}}]{Matevosyan:2018icf}%
  \BibitemOpen
  \bibfield  {author} {\bibinfo {author} {\bibfnamefont {H.~H.}\ \bibnamefont
  {Matevosyan}}, \bibinfo {author} {\bibfnamefont {A.}~\bibnamefont
  {Bacchetta}}, \bibinfo {author} {\bibfnamefont {D.}~\bibnamefont {Boer}},
  \bibinfo {author} {\bibfnamefont {A.}~\bibnamefont {Courtoy}}, \bibinfo
  {author} {\bibfnamefont {A.}~\bibnamefont {Kotzinian}}, \bibinfo {author}
  {\bibfnamefont {M.}~\bibnamefont {Radici}}, \ and\ \bibinfo {author}
  {\bibfnamefont {A.~W.}\ \bibnamefont {Thomas}},\ }\href {\doibase
  10.1103/PhysRevD.97.074019} {\bibfield  {journal} {\bibinfo  {journal} {Phys.
  Rev. D}\ }\textbf {\bibinfo {volume} {97}},\ \bibinfo {pages} {074019}
  (\bibinfo {year} {2018})},\ \Eprint {http://arxiv.org/abs/1802.01578}
  {arXiv:1802.01578 [hep-ph]} \BibitemShut {NoStop}%
\bibitem [{\citenamefont {Courtoy}\ \emph {et~al.}(2012)\citenamefont
  {Courtoy}, \citenamefont {Bacchetta}, \citenamefont {Radici},\ and\
  \citenamefont {Bianconi}}]{Courtoy:2012ry}%
  \BibitemOpen
  \bibfield  {author} {\bibinfo {author} {\bibfnamefont {A.}~\bibnamefont
  {Courtoy}}, \bibinfo {author} {\bibfnamefont {A.}~\bibnamefont {Bacchetta}},
  \bibinfo {author} {\bibfnamefont {M.}~\bibnamefont {Radici}}, \ and\ \bibinfo
  {author} {\bibfnamefont {A.}~\bibnamefont {Bianconi}},\ }\href {\doibase
  10.1103/PhysRevD.85.114023} {\bibfield  {journal} {\bibinfo  {journal} {Phys.
  Rev.}\ }\textbf {\bibinfo {volume} {D85}},\ \bibinfo {pages} {114023}
  (\bibinfo {year} {2012})},\ \Eprint {http://arxiv.org/abs/1202.0323}
  {arXiv:1202.0323 [hep-ph]} \BibitemShut {NoStop}%
%%CITATION = ARXIV:1202.0323;%%
\bibitem [{\citenamefont {Bacchetta}\ \emph {et~al.}(2007)\citenamefont
  {Bacchetta}, \citenamefont {Diehl}, \citenamefont {Goeke}, \citenamefont
  {Metz}, \citenamefont {Mulders} \emph {et~al.}}]{Bacchetta:2006tn}%
  \BibitemOpen
  \bibfield  {author} {\bibinfo {author} {\bibfnamefont {A.}~\bibnamefont
  {Bacchetta}}, \bibinfo {author} {\bibfnamefont {M.}~\bibnamefont {Diehl}},
  \bibinfo {author} {\bibfnamefont {K.}~\bibnamefont {Goeke}}, \bibinfo
  {author} {\bibfnamefont {A.}~\bibnamefont {Metz}}, \bibinfo {author}
  {\bibfnamefont {P.~J.}\ \bibnamefont {Mulders}},  \emph {et~al.},\ }\href
  {\doibase 10.1088/1126-6708/2007/02/093} {\bibfield  {journal} {\bibinfo
  {journal} {JHEP}\ }\textbf {\bibinfo {volume} {0702}},\ \bibinfo {pages}
  {093} (\bibinfo {year} {2007})},\ \Eprint
  {http://arxiv.org/abs/hep-ph/0611265} {arXiv:hep-ph/0611265 [hep-ph]}
  \BibitemShut {NoStop}%
%%CITATION = HEP-PH/0611265;%%
\bibitem [{\citenamefont {Radici}\ \emph {et~al.}(2015)\citenamefont {Radici},
  \citenamefont {Courtoy}, \citenamefont {Bacchetta},\ and\ \citenamefont
  {Guagnelli}}]{Radici:2015mwa}%
  \BibitemOpen
  \bibfield  {author} {\bibinfo {author} {\bibfnamefont {M.}~\bibnamefont
  {Radici}}, \bibinfo {author} {\bibfnamefont {A.}~\bibnamefont {Courtoy}},
  \bibinfo {author} {\bibfnamefont {A.}~\bibnamefont {Bacchetta}}, \ and\
  \bibinfo {author} {\bibfnamefont {M.}~\bibnamefont {Guagnelli}},\ }\href
  {\doibase 10.1007/JHEP05(2015)123} {\bibfield  {journal} {\bibinfo  {journal}
  {JHEP}\ }\textbf {\bibinfo {volume} {05}},\ \bibinfo {pages} {123} (\bibinfo
  {year} {2015})},\ \Eprint {http://arxiv.org/abs/1503.03495} {arXiv:1503.03495
  [hep-ph]} \BibitemShut {NoStop}%
%%CITATION = ARXIV:1503.03495;%%
\bibitem [{\citenamefont {Bacchetta}\ \emph {et~al.}(2013)\citenamefont
  {Bacchetta}, \citenamefont {Courtoy},\ and\ \citenamefont
  {Radici}}]{Bacchetta:2012ty}%
  \BibitemOpen
  \bibfield  {author} {\bibinfo {author} {\bibfnamefont {A.}~\bibnamefont
  {Bacchetta}}, \bibinfo {author} {\bibfnamefont {A.}~\bibnamefont {Courtoy}},
  \ and\ \bibinfo {author} {\bibfnamefont {M.}~\bibnamefont {Radici}},\ }\href
  {\doibase 10.1007/JHEP03(2013)119} {\bibfield  {journal} {\bibinfo  {journal}
  {JHEP}\ }\textbf {\bibinfo {volume} {03}},\ \bibinfo {pages} {119} (\bibinfo
  {year} {2013})},\ \Eprint {http://arxiv.org/abs/1212.3568} {arXiv:1212.3568
  [hep-ph]} \BibitemShut {NoStop}%
%%CITATION = ARXIV:1212.3568;%%
\end{thebibliography}
\end{document}